\documentclass[preprint, prl, superscriptaddress]{revtex4}
\usepackage[dvips]{graphicx}
\newcommand{\futoi}[1]{\mbox{\boldmath$#1$}}

\begin{document}

\title
{
Room Temperature Reversible Spin Hall Effect
}

\author
{ 
T. Kimura 
}
\affiliation{
Institute for Solid State Physics, University of Tokyo 
5-1-5 Kashiwanoha, Kashiwa, Chiba 277-8581, Japan
}
\affiliation{
RIKEN FRS, 2-1 Hirosawa, Wako, Saitama 351-0198, Japan 
}
\email{kimura@issp.u-tokyo.ac.jp}

\author
{
Y. Otani 
}
\affiliation{
Institute for Solid State Physics, University of Tokyo 
5-1-5 Kashiwanoha, Kashiwa, Chiba 277-8581, Japan
}
\affiliation{
RIKEN FRS, 2-1 Hirosawa, Wako, Saitama 351-0198, Japan 
}

\author
{ 
T. Sato
}
\affiliation{
Institute for Solid State Physics, University of Tokyo 
5-1-5 Kashiwanoha, Kashiwa, Chiba 277-8581, Japan
}

\author
{
S. Takahashi
}
\affiliation{
Institute for Materials Research, Tohoku University, Sendai, Miyagi 980-8577, Japan
}
\affiliation{
CREST, JST, Honcho 4-1-8, Kawaguchi, Saitama, 332-0012, Japan 
}

\author
{
S. Maekawa
}
\affiliation{
Institute for Materials Research, Tohoku University, Sendai, Miyagi 980-8577, Japan
}
\affiliation{
CREST, JST, Honcho 4-1-8, Kawaguchi, Saitama, 332-0012, Japan 
}

\date{\today}
\begin{abstract}
Reversible spin Hall effect comprising the "direct" and "inverse" 
spin Hall effects was successfully detected at room temperature. 
This experimental demonstration proves the fundamental relations called Onsager reciprocal 
relations between spin and charge currents. 
A platinum wire with a strong spin-orbit interaction is used not only as a spin current 
absorber but also as a spin current source in the present lateral structure specially designed 
for clear detection of both charge and spin accumulations via the spin-orbit interaction. 
The obtained spin Hall conductivity is much larger than the reported value of Aluminum wire 
because of the larger spin-orbit interaction.  

\end{abstract}

\maketitle

The basic science for electronic devices aiming at manipulating the spin degree of 
freedom is spintronics, which provides a possible means to realize advantageous functionalities 
for spin based recording and information processing.  
For such functions, the usage of spin current, a flow of spin angular momentum, is indispensable.  
Thus establishing techniques for efficient generation and manipulation of spin currents 
is a key for further advancement of spintronic devices.  
There is a novel phenomenon where the spin-orbit interaction converts 
a charge current into a spin current and vice versa \cite{SHE1, SHE2, SHE3}.  
These are known as the "direct" and "inverse" spin Hall effects (SHEs). 
 
The SHE was first predicted in 1971 by D'yakonov and Perel'\cite{SHE1}, 
followed by the phenomenological theory with impurity scattering developed by Hirsch \cite{SHE2}, 
and extended to the diffusive transport regime by Zhang\cite{SHE3}. 
Recently, the SHE has attracted considerable interest because of the potentiality for generation 
and manipulation of spin currents in nonmagnets without external magnetic fields. 
Although there have recently been many theoretical studies on the SHE 
(\cite{SHE1,SHE2, SHE3, SHE4, SHE5, SHE6, SHE7, 
SHE8, SHE9, SHE10}), 
few experimental reports exist because of difficult sample fabrication and measurements. 
Kato et al. reported the first experimental observation of the SHE induced spin accumulation 
via magneto-optical Kerr effect in GaAs semiconductor systems\cite{SHE11}.  
Shortly after this measurement, the SHE in a 2 DEG system was demonstrated 
using similar optical technique\cite{SHE12}.  
Although two kinds of extrinsic \cite{SHE1, SHE2, SHE3} and intrinsic \cite{SHE4, SHE5} mechanisms 
for the SHE are proposed to explain the experimental results\cite{SHE11, SHE12, SHE13}, 
the detailed origin of the observed phenomena is still unclear.

In diffusive normal metals, the SHE is known to be induced by the spin-orbit scattering, 
which is an extrinsic effect due to impurities or defects \cite{SHE3, SHE7, SHE8, SHE9, SHE10}. 
Since the optical detection technique is limited for semiconductor systems, the electrical 
detection technique is suitable for observing the SHE in diffusive metals. 
The unpolarized charge current flowing in a nonmagnet generates the transverse spin current, 
and results in the spin accumulation along the side edge of the nonmagnet. 
Inversely, the spin current flowing in a nonmagnet induces the transverse charge current, 
and causes the charge accumulation (inverse SHE)\cite{SHE6, SHE10}. 
Very recently, the first clear observation of the charge accumulation due to the inverse SHE 
is reported by using a nonlocal spin injection 
in a lateral ferromagnetic/nonmagnetic metal structure\cite{SHE14}. 
However, the observation has been performed only at low temperatures possibly because 
of a small spin-orbit interaction of aluminium.  
To induce a large measurable inverse SHE even at room temperature, 
a large spin-orbit scattering is required. 
Platinum is known to exhibit a large spin-orbit interaction because of the large atomic number 
with respect to nonmagnetic impurities or defects\cite{SHE15}.  
However, the conventional lateral structure \cite{SHE14, SHE16} with nonlocal spin injection 
is not applicable to detect the SHE because of 
an extremely short spin diffusion length of about 10 nm \cite{SHE15}. 
There is an experimental report of the charge accumulation due to the inverse SHE 
in the Pt strip using a spin pumping technique\cite{SHE17}. 
However, the large sample size makes it difficult to evaluate the important parameters related to the SHE. 

Although the "inverse" SHE has been reported by the groups mentioned above, 
there has never been electrically observed the "direct" SHE 
(, i. e. , the conversion of the charge current to the spin current,) 
in all metallic systems up to now. 
Here we report a clear observation of the spin and charge accumulations due to the direct 
and inverse SHEs at room temperature induced in the Pt wire using nonlocal spin injection, 
spin absorption, and spin-generation techniques. 
We also demonstrate the Onsager reciprocal relations between the "direct" and the "inverse" SHEs. 
  
Our device for the present SHE experiment consists of lateral ferromagnetic/nonmagnetic metallic 
junctions as shown in Fig.\ 1(a). 
The device consists of a large Permalloy (Py) pad 30 nm in thickness, 
a Cu cross 100 nm in width and 80 nm in thickness, and a Pt wire 80 nm 
in width and 4 nm in thickness. 
The size of the junction between the Py injector and Cu wire, 
is chosen to be 100 nm $\times$ 100 nm 
to induce a large spin accumulation in the Cu wire. 
The distance from the center of the injector to the center of the Pt wire is 400 nm. 
The resistivities of the Py, Cu and Pt are $15.4\times 10^{-8}$ $\Omega$m, 
$2.1\times 10^{-8}$ $\Omega$m and $15.6\times 10^{-8}$ $\Omega$m at RT and 
$10.2 \times 10^{-8}$ $\Omega$m,  $1.0 \times 10^{-8}$ $\Omega$m
and $12.8\times 10^{-8}$ $\Omega$m at 77K, respectively. 
The surfaces of Py injector and Pt wire is carefully cleaned 
by low voltage Ar ion milling prior to the Cu deposition. 
The obtained interfaces have very low resistances in the milli-ohm range, 
indicating good ohmic contact.

When the charge current is injected from a ferromagnetic Py pad into a non-magnetic Cu cross 
and is drained from one of the top and bottom arms (see inset of Fig.\ 2(b)), the accumulated spins 
at the junction induce the diffusive flow of the spin current along the Cu wire as can be understood 
from the electrochemical potential map in Fig.\ 1(b). 
When the distance between the Py/Cu and Cu/Pt junctions $d$ is shorter than the spin diffusion length 
of about 500 nm\cite{SHE18}, the spin current should be absorbed into the Pt wire from the Cu cross.  
According to previous experiments\cite{SHE19}, the magnitudes of the spin current and spin accumulation 
can be calculated by taking into account the spin resistance, defined as $R_S = \lambda / (\sigma S (1-P^2))$ with 
the spin diffusion length $\lambda$, the spin polarization $P$, the conductivity $\sigma$, 
and the effective cross sectional area $S$ for the spin current. 
The physical meaning of the spin resistance is a measure of the difficulty for spin 
mixing over the spin diffusion length. 
The spin current in the spin-accumulated material is thus preferably absorbed 
into the additionally connected material with small spin resistance. 
Here the spin resistance of Pt wire is an order of magnitude smaller than that of Cu wire. 
Therefore, the induced spin current in the Cu cross can be directed into the Pt wire as shown in Fig.\ 1(c). 
  
The injected spin current into the Pt wire 
vanishes immediately in the vicinity of the Pt/Cu junction 
because of the short spin diffusion length of the Pt. 
Hence, the spin current flows almost perpendicular to the junction plane 
as in Fig.\ 1(c)\cite{SHE19,SHE20}. 
The charge current $\futoi{I_C}$ is generated in the Pt wire via the inverse SHE 
when the spin current $\futoi{I_S}$ enters the Pt wire. 
The direction of the current $\futoi{I_C}$ is given by the vector product $\futoi{s} \times \futoi{I_S}$, 
where $\futoi{s}$ is the spin direction, yielding the flow of charge current normal 
to both the spin current $\futoi{I_S}$ and the spin direction $\futoi{s}$. 
As shown in Figs.\ 1(b) and 1(c), 
the up- and down-spin electrons flow opposite to each other as the flow of the electron is 
determined by the gradient of electrochemical potential. 
Such flows of up- and down-spins could induce the transverse voltage due to the spin-orbit interaction. 
Thus, when the spin currents polarized along the x-axis is injected into the Pt wire, 
the charge accumulation along the y-axis is induced in the Pt wire. 
It should be remarked here that the inverse transformation from the charge current $\futoi{I_C}$ 
to the spin current $\futoi{I_S}$ is also expected when the configuration of injector and detector probes 
is reversed as illustrated in Figs.\ 1(d) and 1(e). 

In order to observe the charge accumulation due to the inverse SHE, 
we first measure the Hall voltage $V_C$ 
induced in the Pt wire by means of the nonlocal spin injection. 
The spin-polarized charge current is injected from the Py pad into the Cu cross. 
The magnetic field is applied along the x-axis to maximize the charge accumulation in the Pt wire. 
As shown in Figs.\ 2(a) and 2(b), the clear change in $\Delta V_C /I$ appears both at RT and 77 K. 
The $\Delta V_C /I$ curves show hysteresis, assuring that the charge accumulation is induced by the spin current 
from the Py injector. The magnitude of $\Delta R_{\rm SHE}$, defined 
as the overall change of $\Delta V_C /I$, is increased up to 160 m$\Omega$ at 77K. 
The observed change in signal does not depend on the choice of the Cu arm 
and the similar results were observed in other two devices, assuring good reproducibility.  
One should notice that in this device configuration the charge accumulation can be induced 
by the in-plane component of the Py magnetization. 
This is a great advantage because the influence of the demagnetizing field can be avoided. 
Furthermore unlike previous metallic device the magnetization direction of the spin injector Py 
can be controlled at will by small in-plane magnetic fields below a few hundreds Oersted. 
  
To understand the relation between the direction of the injected spins $\futoi{s}$ 
and the induced charge accumulation, 
the angular dependence of $\Delta V_C /I$ is measured as a function of the in-plane field H 
along the tilting angle f from the x-axis. The induced $\Delta R_{\rm SHE}$ decreases 
with transverse magnetic component (Figs.\ 3(a), and 3(b)), 
and vanishes when the magnetization is aligned with the y-axis (Fig.\ 3(c)). 
From the relation $\futoi{I_{C}} \propto \futoi{s} \times \futoi{I_S}$, 
the angular dependence of the $\Delta R_{\rm SHE}$ is expected to vary with $\cos \phi$. 
Figure 3(d) shows clear $\cos \phi$ variations of $\Delta R_{\rm SHE}$ both at RT and 77 K 
in good agreement with the prediction. 
  
Finally, the transformation from the charge current $\futoi{I_C}$ to the spin current $\futoi{I_S}$, 
which is the "direct" SHE, is examined using the probe configuration in the inset of Fig.\ 4(b). 
It should be noted that in the above described experiment, the Pt wire is used as a spin-current absorber, 
but the Pt wire now acts as a spin-current source, the latter of which is much more important 
for the application of spintronics without using a ferromagnet. 
Figures 4(a) and 4(b) show the field dependences of the induced spin accumulation signal 
$\Delta V_S / I$ in the Pt wire  measured at RT and 77 K. 
$\Delta V_S / I$ varies similarly to $\Delta V_C/I$ in Figs.\ 2(a) and 2(b) 
and more importantly the overall resistance change $\Delta R_S$ is exactly the same as $\Delta R_{\rm SHE}$. 
These results demonstrate that both the spin to charge and the charge to spin current transformations 
are reversible, corresponding to the Onsager reciprocal relations, 
$\sigma_{\rm SHE} = \sigma'_{\rm SHE}$, 
where $\sigma_{\rm SHE}$ and $\sigma'_{\rm SHE}$ are respectively 
the spin-current-induced spin-Hall conductivity and the charge-current-induced spin-Hall conductivity\cite{SHE4}. 
The spin Hall conductivity of the Pt wire is here evaluated as follows; 
In the experiments of spin-current induced charge accumulation shown 
in Figs.\ 1(b) and 1(c), the injected spin current $I_{Si}$ into the Pt wire is approximated by assuming 
that both the spin resistances $R_{S{\rm Py}}$ of Py and $R_{S{\rm Pt}}$ of Pt 
are much smaller than $R_{S{\rm Cu}}$ of Cu\cite{SHE10,SHE19, SHE20}:
\begin{equation}
I_{Si} \approx P_{\rm Py} I_C \frac{R_{S \rm Py}}{R_{S \rm Cu}} 
\left( \sinh \left(\frac{d}{\lambda_{\rm Cu}} \right) \right)^{-1}
\end{equation}
where $\lambda_{\rm Cu}$, $P_{\rm Py}$ and $I_{C}$ are the spin diffusion length of Cu, 
the spin polarization of Py and the injected charge current. 
The spin Hall conductivity $\sigma_{\rm SHE}$ can be given by the following equation\cite{SHE3, SHE8,SHE14}:  
\begin{equation}
\sigma_{\rm SHE} \approx 
w \sigma_{\rm Pt}^2 \Delta R_{\rm SHE} \left( \frac{I_C}{I_{Si}} \right) 
= \left( \frac{w \sigma_Pt^2}{P_{\rm Py}} \right)
\left( \frac{R_{S \rm{Cu}}}{R_{S \rm{Py}}}  \right) 
\sinh \left(\frac{d}{\lambda_{\rm Cu}} \right)
\Delta R_{\rm SHE}
\end{equation}
where $w$ is the width of the Pt wire (80 nm in the present study). 
By using $P_{\rm Py\ at\ RT}$  = 0.2, $\lambda_{\rm Py\ at\  RT}$ = 3 nm, 
$\lambda_{\rm Cu\ at\ RT}$ = 500 nm, we obtain the value of $\sigma_{\rm SHE}$ as 
2.4 $\times 10^4$ ($\Omega$m)$^{-1}$. 
Thus, the ratio of the spin Hall conductivity to the electrical conductivity 
$\alpha_{\rm SHE} = \sigma_{\rm SHE}/\sigma$ for Pt is deduced to be 3.7 $\times$ 10$^{-3}$, 
30 times larger than that for Al obtained in a previous experiment\cite{SHE14}, 
as expected from the larger atomic number of Pt compared with Al. 
The dimensionless spin orbit parameter $\eta_0$ can also be calculated to be 0.74 
from the relation  with the Fermi wave vector $k_{F} \approx$ 1 $\times$ $10^{10}$ m$^{-1}$
and the mean free path $l \approx$ 20 nm for Pt\cite{SHE15}. 
This is quantitatively in good agreement with the value of 0.80 separately calculated using the equation 
$\eta_0 = (3 \sqrt{3} \pi/2)(R_K/k_F^2)(\sigma/\lambda)$ with the quantum resistance $R_K$\cite{SHE10}. 
This result opens up a new possibility to use normal metals with high spin-orbit coupling 
as spin current sources operating at room temperature for the future spintronic applications.

\section*{Acknowledgements}
We thank Prof. Y. Iye and Prof. S. Katsumoto of ISSP, Univ. of Tokyo for the use of the lithography facilities.

\begin{figure}

\caption{
(a) Scanning electron microscope (SEM) image of the fabricated spin Hall 
device together with a schematic illustration of the fabricated device. 
(b) Schematic spin dependent electrochemical 
potential map indicating spin accumulation in Cu and Pt induced 
by the spin injection from the Py pad. Dashed line represents 
the equilibrium position. 
The charge current is injected from the Py pad into the Cu cross, 
and is drained from one of the Cu arms. At the junction 3, there 
is no charge current. The Pt/Cu junction acts as a strong spin absorber (spin sink) 
because of the smallest spin resistance of Pt in the system. 
(c) Schematic illustration of the charge accumulation process in the Pt wire, 
where $\futoi{I_S}$ and $\futoi{I_e}$ denote injected pure spin current 
and induced charge current, respectively. 
At the junction between the Cu and Pt wires, the spin current 
$\futoi{I_S}$ perpendicularly flows into the junction plane. 
(d) Spin dependent electrochemical potential map for the charge to spin current conversion and 
(e) corresponding schematic illustration. 
The spin current generated in the Pt wire flows along the z-axis into the Cu wire.
}

\caption{
The change in Hall resistance $\Delta V_C/I$ due to the inverse spin Hall effect (SHE) 
(a) at room temperature and (b) at 77K. The inset shows the probe configuration for the measurement. 
The charge current is injected from the Py pad into the Cu cross and is drained from the upper Cu arm. 
The external magnetic field is applied along the x-axis to maximize charge accumulation along the Pt wire. 
The measurements are performed by using a current-bias lock-in technique with the excitation current of 300 mA.
$\Delta R_{\rm SHE}$ is defined as the overall change from the maximum to minimum.  
}

\caption{
Hall resistance change due to the inverse SHE at room temperature for various directions 
((a) $\phi = 0$, (b) $\phi = \pi/4$ and (c) $\phi = \pi$) 
of the external magnetic field. 
(d) Overall Hall resistance change $\Delta R_{\rm SHE}$ 
as a function of the direction of the magnetic field.  
The solid lines are $\cos \phi$ 
curves best fitted to the results.
}

\caption{
Spin accumulation signal $\Delta V_{\rm S}/I$  generated by SHE at room temperature and 77 K.  
Field dependences of the signal are similar to those in Figs. 2(a) and 2(b). 
The overall change from maximum to minimum is the same as that in Fig. 2. 
This is consistent with the theoretical prediction. 
The exciting current for the lock-in measurements is limited below 60 mA 
to prevent Joule heating of the thin Pt wire 4 nm in thickness. 
To obtain reasonable S/N ratio in the signal for the room temperature measurement, 
30 curves are averaged.  
}

\end{figure}

\newpage
\vspace*{4cm}
\begin{center}
\includegraphics[width=8cm]{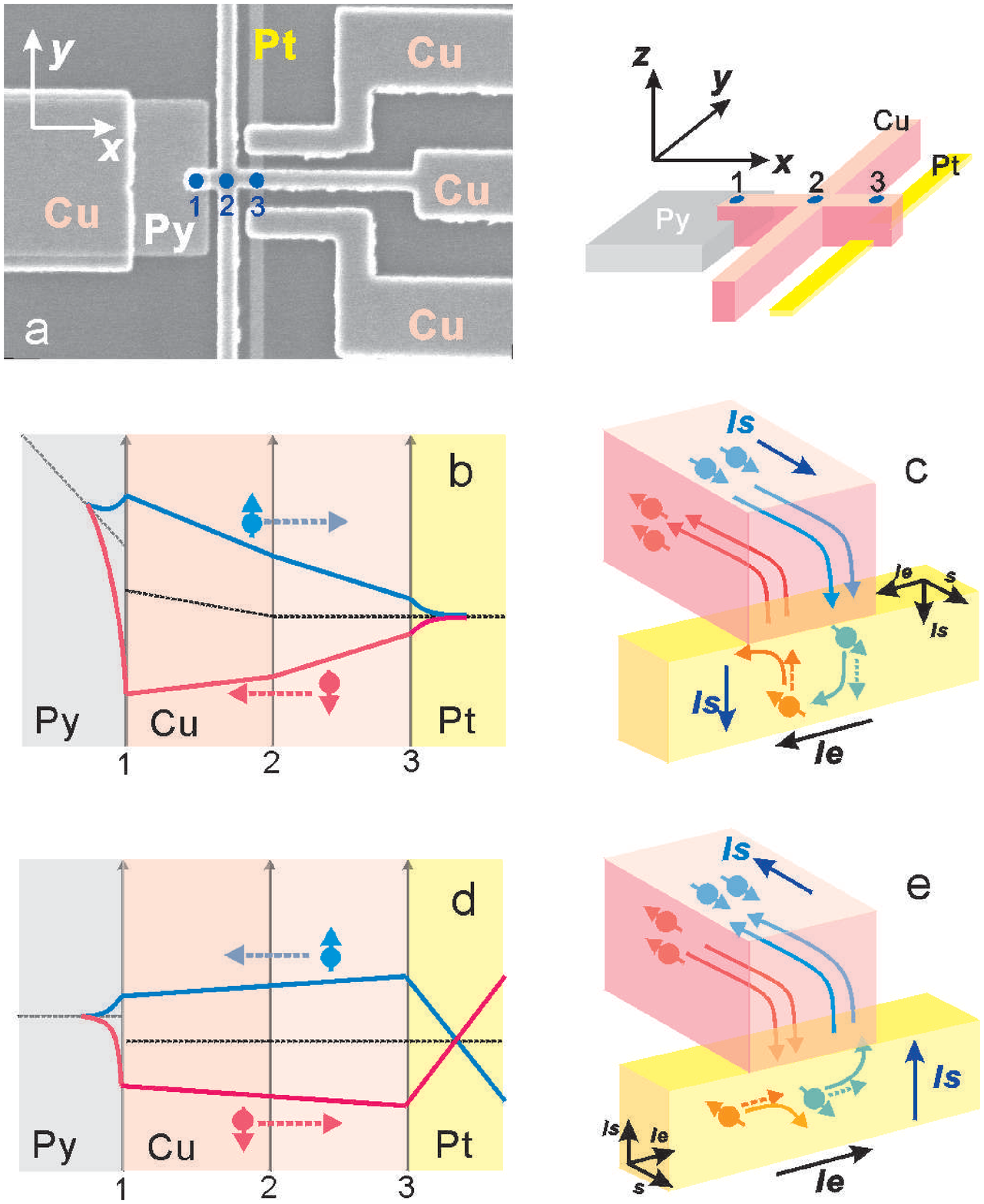}
\end{center}
\vspace*{1cm}
\begin{center}
Fig.\ 1 Kimura et al.
\end{center}

\newpage
\vspace*{4cm}
\begin{center}
\includegraphics[width=8cm]{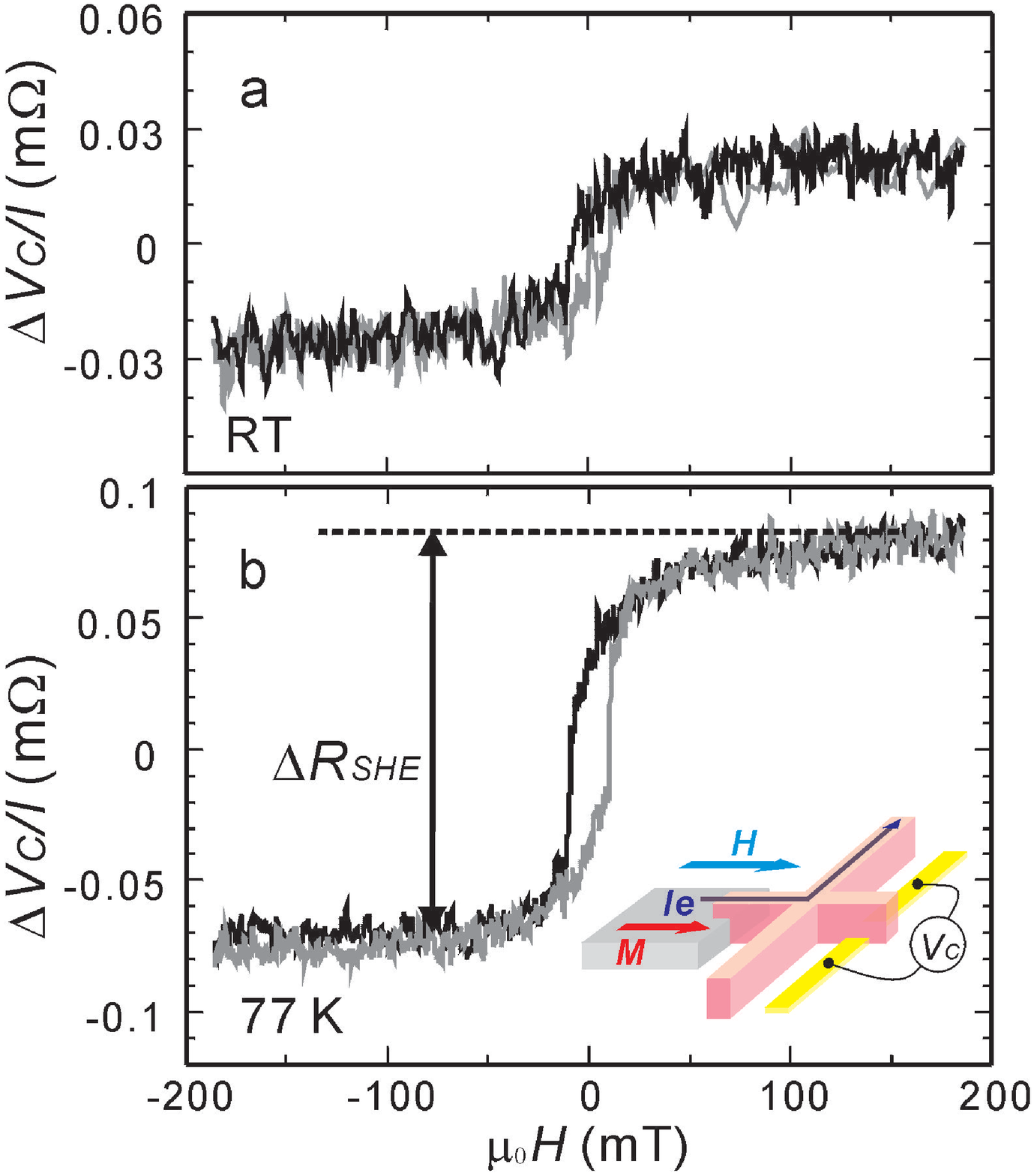}
\end{center}
\vspace*{1cm}
\begin{center}
Fig.\ 2 Kimura et al.
\end{center}

\newpage
\vspace*{4cm}
\begin{center}
\includegraphics[width=8cm]{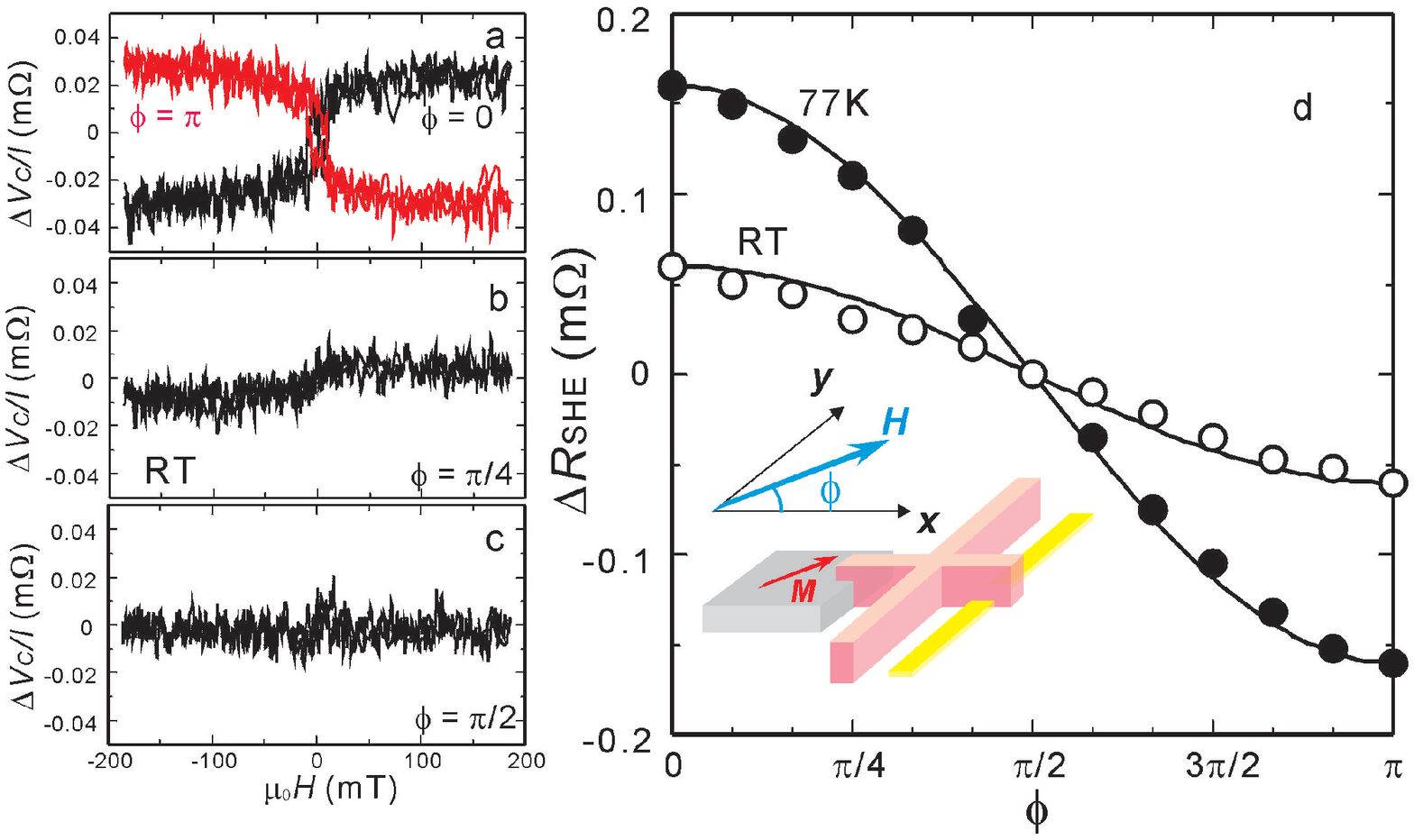}
\end{center}
\vspace*{1cm}
\begin{center}
Fig.\ 3 Kimura et al.
\end{center}

\newpage
\vspace*{4cm}
\begin{center}
\includegraphics[width=8cm]{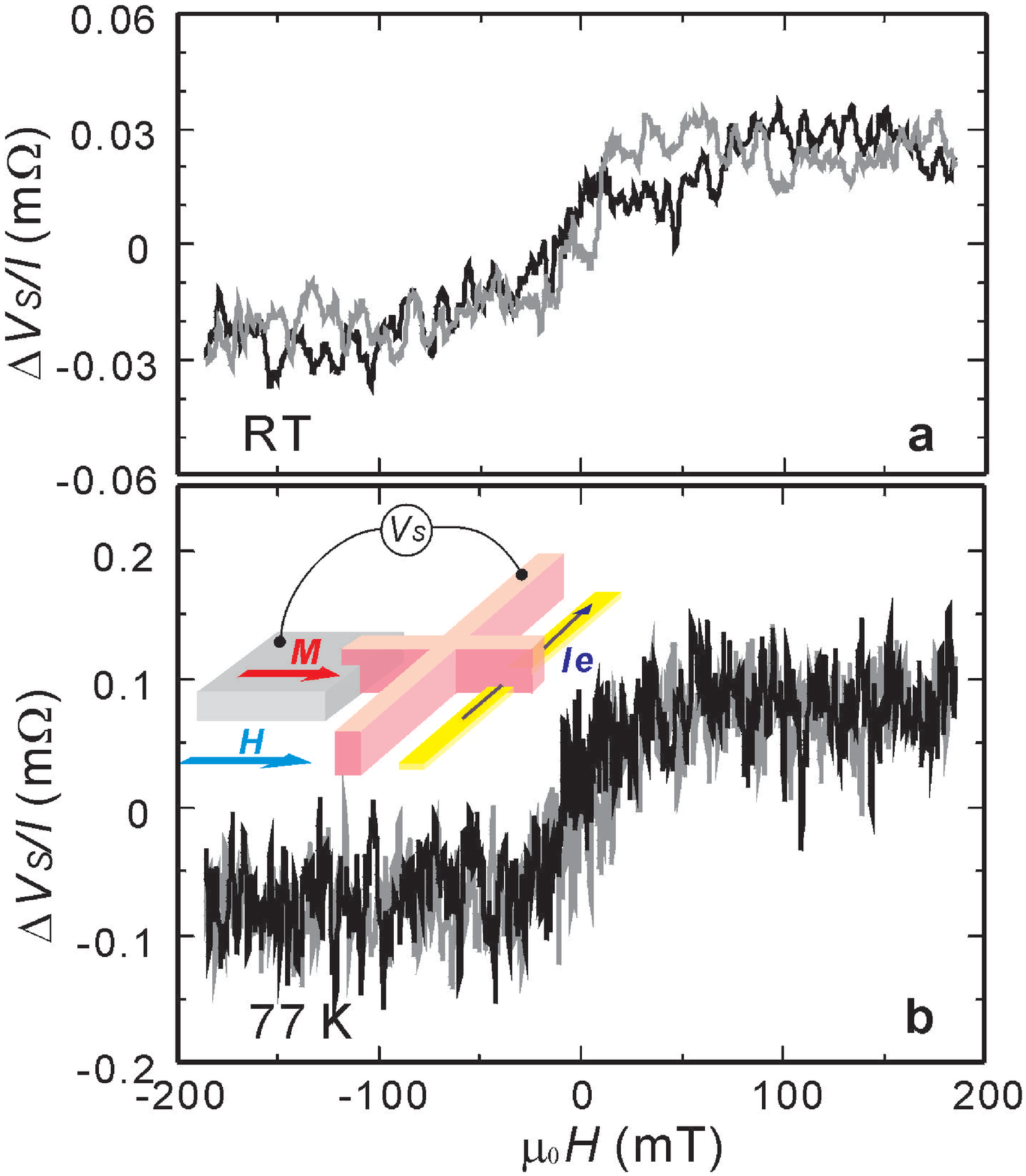}
\end{center}
\vspace*{1cm}
\begin{center}
Fig.\ 4 Kimura et al.
\end{center}

\end{document}